\documentclass[preprint,showpacs,preprintnumbers,amsmath,amssymb,superscriptaddress]{revtex4}
\usepackage{graphicx}
\usepackage{bm}
\usepackage{color,graphics}
\definecolor{orange}{rgb}{.95,.75,0}
\begin{document}
\def\mstar{m_{\displaystyle *}}
\def\revise{\color{blue}}
\title{Physical interpretation of the quantum two-stream instability}
\author{F. Haas}
\affiliation{Institut f\"ur Theoretische Physik IV,
Ruhr-Universit\"at Bochum, D-44780 Bochum, Germany}
\author{A. Bret}
\affiliation{ETSI Industriales, Universidad de Castilla-La Mancha, 13071 Ciudad Real, Spain}
\author{P. K. Shukla}
\affiliation{Institut f\"ur Theoretische Physik IV,
Ruhr-Universit\"at Bochum, D-44780 Bochum, Germany}
\begin{abstract}
\noindent
The unexpected features of the two-stream instability in electrostatic quantum plasmas are interpreted in terms of the coupling of approximate fast and slow waves. This is accomplished through the factorization of the dispersion relation into different sectors having positive or negative energy. Therefore, the concept of negative and positive energy waves is useful not only for classical, but for quantum plasmas as well. We discuss the limitations of the quantum two-stream model in view of the weak coupling assumption.
\end{abstract}


\maketitle

\section{Introduction}
Since it has been discussed for the first time in the framework of the quantum hydrodynamical model \cite{Haas}, the quantum two-stream instability has attracted considerable attention in the literature. The reason for this is that it is a benchmark displaying many of the par\-ti\-cu\-la\-ri\-ties of quantum plasmas, including a new unstable branch of the dispersion relation for large wave-numbers and almost stationary, quasineutral, nonlinear oscillations \cite{Haas} without analog in classical plasmas. Later on, a kinetic (Wigner-Poisson) treatment of the problem \cite{Anderson} showed that temperature effects can suppress the quantum instability. The quantum fluid equations have been further applied to several quantum streaming instability problems, like the three-stream quantum plasmas \cite{Haas2},  the quantum dusty plasmas \cite{Ali},  the electron-positron-ion quantum plasmas \cite{Mushtaq} or the magnetized multi-stream instability \cite{Ren}. The hydrodynamic formalism has also been applied to the quantum filamentation instability, with \cite{Bret} or without \cite{Bret2} magnetization.

The purpose of the present contribution is to provide an intuitive explanation of the quantum two-stream instability, in terms of the coupling of electrostatic modes with distinct energy contents. Indeed, literature  still  lacks  a physical (``with the hands") understanding of the quantum streaming instabilities. In addition to analytic and numeric approaches, the physical interpretation of quantum plasma effects is a most welcome extension, helping to describe the unexpected features of these problems. As will be shown in the sequel, it is possible to identify approximate positive and negative energy modes in electrostatic two-stream quantum plasmas. Moreover, the energy exchange between these waves provide the clue for the stability analysis in such systems.

In addition, we also examine the range of validity of the quantum two-stream model. As shown in Section IV, the quantum parameter $H$, which is the ratio between the plasmon energy and the kinetic energy of each beam, should be small in order to fulfill the collisionless assumption. As far as we know, a clear statement about the smallness of $H$ is missing. Nevertheless, the more intriguing aspects of the quantum two-stream instability appear for small $H$, where the so-called quantum instability branch for large wave-numbers appear \cite{Haas}. The present contribution gives a physical justification for this exotic unstable quantum branch in terms of negative energy waves. Finally, the fact that a large quantum parameter tends to enhance collisions is a reason more to  develop a (still lacking) efficient  collisional quantum plasma theory.

Negative energy modes are a well-known tool for the analysis of streaming instabilities in classical systems \cite{Sturrock, Akhiezer}. In this work we show that the same ideas are also useful to quantum plasmas. The basic heuristic concept of negative energy wave is as follows. Consider, for definiteness, coherent wave motion taking place in some dispersive medium crossed by a number of streaming particles. When the absolute value of the wave's phase velocity is slightly smaller than the speed of some of the other streams, the oscillation mode is referred to as a negative energy wave. The mechanism of the classical two-stream instability can be explained in terms of energy transfer from a positive energy Langmuir wave taking place in one beam,  to a negative energy Langmuir wave taking place in the second beam. Naturally in quantum plasma physics such a classical picture of resonant wave--particle interaction should be taken with caution. However, we will show that negative energy modes can still be identified in terms of macroscopic (fluid) quantum plasma equations for the two-stream instability.

This work is organized as follow. In Section II, we review the basic features of the quantum two-stream instability, and of the electric field energy density in a weakly dissipative dielectric medium. In Section III, fast and slow approximate waves in two-stream quantum plasmas are identified. The coupling of these waves is then analyzed in detail, in order to understand why some wave-numbers are stable and others are not, when varying the strength of quantum effects. A detailed account on the limitations of the quantum two-stream model for large quantum parameters $H$ due to the neglect of collisions is included in Section IV. Section V is reserved to the conclusions.

\section{Time-averaged energy density of electrostatic oscillations}
Consider two counter propagating electron beams with equal equilibrium number densities $n_1 = n_2 = n_0/2$, and equilibrium velocities $u_1 = - u_2 = u_0 > 0$, in presence of an immobile ionic background of particle density $n_0$. Following the same notation of Ref. \cite{Haas}, it is convenient to normalize the frequencies to the plasma frequency $\omega_p$ and the wave-numbers to $\omega_p/u_0$. Using the two-stream quantum hydrodynamic model \cite{Haas}, a dielectric function $\varepsilon = 1 - F(\Omega)$ is derived, where the characteristic function $F$ is
\begin{equation}
\label{e1}
F(\Omega) =  \frac{1}{2}\left[\frac{1}{(\Omega+K)^2- H^2\,K^4/4} + \frac{1}{(\Omega-K)^2- H^2\,K^4/4}\right] \,,
\end{equation}
in terms of a non-dimensional quantum parameter $H = \hbar\omega_{p}/mu_{0}^2$, where $\hbar$ is Planck's constant over $2\pi$ and $m$ is the electron mass.

The dispersion relation $\varepsilon = 0$ is a second-order polynomial equation for  $\Omega^2$, with solutions $\Omega^2 = \Omega_{\pm}^{2}(K)$, where
\begin{eqnarray}
\label{e2}
\Omega_+ &=& \frac{1}{2}\left[2+4K^2+H^2\,K^4 + 2\sqrt{1+8K^2+4H^2\,K^6}\,\right]^{1/2} \,,\\
\label{e3}
\Omega_- &=& \frac{1}{2}\left[2+4K^2+H^2\,K^4 - 2\sqrt{1+8K^2+4H^2\,K^6}\,\right]^{1/2} \,,
\end{eqnarray}
corresponding to four possible branches of the eigen-frequency $\Omega$ as a function of the wave-number $K$. We use parity properties to restrict the analysis to positive $K$ and $\Omega$ values. As detailed in Ref. \cite{Haas}, when $0 < H < 1$ there is instability provided $K < K_A$ (semiclassical branch) or $K_B < K < K_C$ (quantum branch), where $K_A < K_B < K_C$ are given by
\begin{equation}
\label{e4}
K_A = \frac{\left[2-2\sqrt{1-H^2}\,\right]^{1/2}}{H} \,,\quad K_B = \frac{\left[2+2\sqrt{1-H^2}\,\right]^{1/2}}{H} \,,\quad K_C = \frac{2}{H} \,.
\end{equation}
On the other hand, when $H \ge 1$ the instability condition is just $K < K_C$ (see Fig. 1 of Ref. \cite{Haas}).

In order to evaluate the  energy content of each wave mode, it is necessary to consider the time-averaged energy density $<W_{e}>$ for   longitudinal plasma oscillations, which is
\begin{equation}
\label{e5}
<W_e> = \frac{\varepsilon_0}{4}\,\frac{\partial(\Omega \,\varepsilon_{h})}{\partial\Omega}\,|E_1|^2 \,,
\end{equation}
where $\varepsilon_0$ is the vacuum permittivity, $\varepsilon_h$ is the Hermitian part of the dielectric function and $E_1$ is the amplitude of the electric field perturbation. Since the underlying model is dissipation-free, one has $\varepsilon = \varepsilon_h$. Eq. (\ref{e5}) is still valid in our quantum plasma system, since it is derived from the Maxwell's equations only. Quantum effects are contained in the modified dielectric function.  Moreover, $<W_e>$ includes both the contributions due to the electrostatic energy and the ``acoustic energy", defined as the kinetic energy associated to the coherent particle wave motion \cite{Stix}.  The expression for $<W_e>$ can equally be deduced from a generalized Poynting theorem, like for the classical two-stream instability \cite{LashmoreDavies}.

Proceeding from Eqs. (\ref{e1}) and (\ref{e5}), one get $\partial(\Omega \,\epsilon_{h})/\partial\Omega \sim \psi(\Omega)$, omitting a complicated positive factor, where
\begin{equation}
\label{e6}
\psi(\Omega) \equiv - 6\, K^4 + H^2\, K^6 + \frac{H^2\, K^8}{8} + K^2 \,(4-H^2 K^2)\,\Omega^2 + 2\, \Omega^4 \,.
\end{equation}
Considering $\Omega = \Omega_{+}(K)$ from Eq. (\ref{e2}), one finds,
\begin{equation}
\label{e7}
\psi(\Omega_{+}) = 1 + 8\,K^2 + 4\,H^2\,K^6 + (1+4\,K^2)\,\sqrt{1+8\,K^2+4\,H^2\,K^6} \, > 0 \,,
\end{equation}
so that this mode is always a {\it positive energy} wave.

On the other hand, taking $\Omega = \Omega_{-}(K)$ from Eq. (\ref{e3}) gives,
\begin{equation}
\label{e8}
\psi(\Omega_{-}) = 1 + 8\,K^2 + 4\,H^2\,K^6 - (1+4\,K^2)\,\sqrt{1+8\,K^2+4\,H^2\,K^6} \,.
\end{equation}
From Eqs. (\ref{e4}) and (\ref{e8}), it can be shown that $\psi(\Omega_{-}) < 0$ if and only if $K < K_C$. Therefore, if $K > K_C$, the mode $\Omega = \Omega_{-}(K)$ is a stable {\it positive energy} mode; if $K < K_C$, the unstable modes with $K_B<K<K_C$ (see Eq. \ref{e4}) and  $\Omega = \Omega_{-}(K)$ are {\it negative energy} waves. Actually, they correspond to an absolute instability as they are of the form $\Omega = i\gamma$, for real $\gamma > 0$. The stable mode $\Omega = \Omega_{-}(K)$ for $K_A < K < K_B$, which exists only for $H < 1$, has negative energy. Finally, one has $\psi(\Omega_{-}) = 0$ for the  marginally stable wave-number $K = K_C$.

Further insight can be gained analyzing the characteristic function $F(\Omega)$ given by Eq. (\ref{e1}). It has vertical asymptotes at $\Omega = \pm \Omega_{>}$ and $\Omega = \pm \Omega_{<}$, where $\Omega_{>} = K + H\,K^2/2$ and $\Omega_{<} = K - H\,K^2/2$. Since the dispersion relation is  quadratic  with real coefficients for $\Omega^2$, stability is assured when the graph of $F(\Omega)$ intercepts the horizontal line $F = 1$ four times. Actually, the case $K = K_C$ is special because the quartic equation for $\Omega$ degenerates into a quadratic one, which can be shown to correspond always to stable oscillations. Figure 1 shows a typical unstable case when $K < K_A$, for $K = 0.9, H = 0.1$.

\begin{figure*}
\begin{center}
\includegraphics[width=0.7\textwidth]{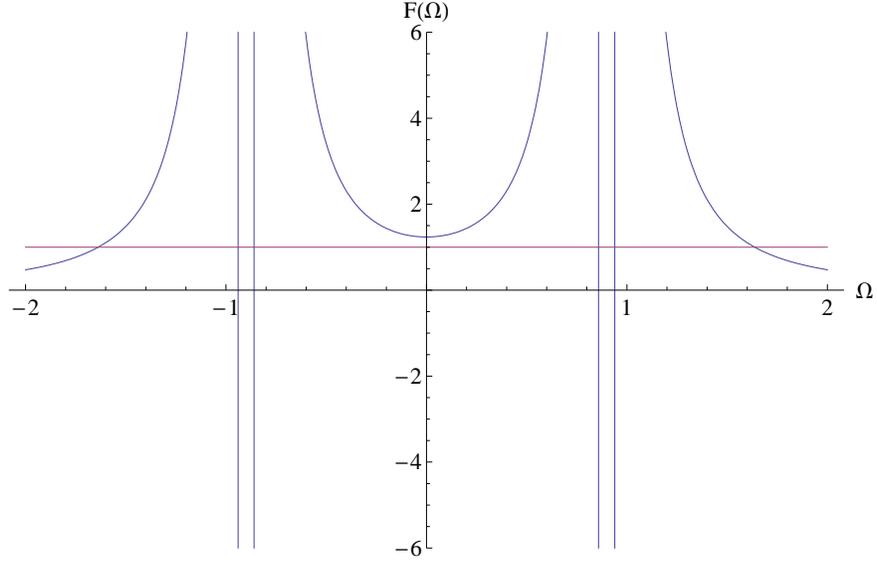}
\caption{Characteristic function for $K < K_A$. In the example, $K = 0.9$, $H = 0.1$. It correspond to (absolute) instability, since $F(0) > 1$.}
\label{figure1}
\end{center}
\end{figure*}

On the other hand, Figure 2 explains why the wave-numbers satisfying $K > K_C$ are stable, since the graph of the characteristic function always intercepts the horizontal line $F = 1$ four times.

\begin{figure*}
\begin{center}
\includegraphics[width=0.7\textwidth]{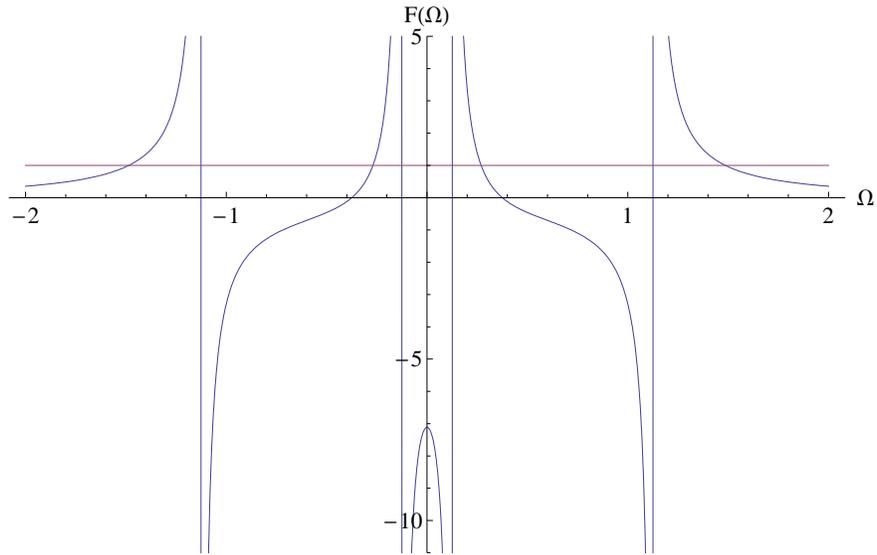}
\caption{Characteristic function for the stable wave-numbers $K > K_C$. In the example, $K = 1$, $H = 0.5$.}
\label{figure2}
\end{center}
\end{figure*}

\section{Fast and slow approximate modes in electrostatic two-stream quantum plasmas}
Figure 3 shows the dispersion curves for $H < 1$. Since the dispersion relation is an algebraic equation with real coefficients, it is possible to follow Sturrock rules \cite{Sturrock, Akhiezer}, identifying several stable/unstable zones in the $(\Omega, K)$ space. Here we consider real wave-numbers and don't analyze the amplification problem. Curve 1 is a positive energy mode parametrized by $\Omega = \Omega_{+}(K)$ given by Eq. (\ref{e2}). Curves 2 and 3 are both described by $\Omega = \Omega_{-}(K)$ given by Eq. (\ref{e3}). However, according to the preceding analysis, curve 2 is a negative energy mode, while curve 3 is a positive energy mode. The coupling of these waves gives rise to the purely quantum (absolute) instability for large wave-numbers, $K_B < K < K_C$ in Fig. 3.

\begin{figure*}
\begin{center}
\includegraphics[width=0.7\textwidth]{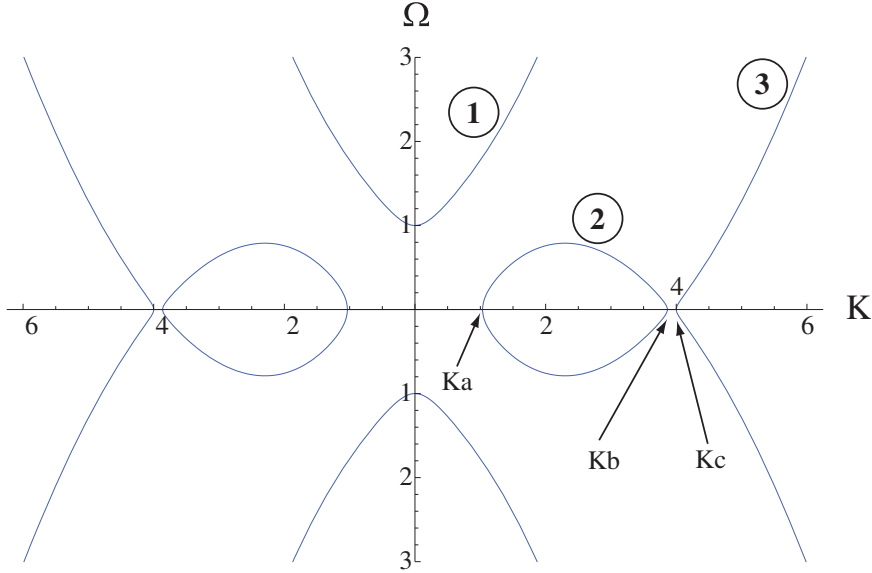}
\caption{Dispersion curves for $H = 0.5$. Curve 1 is a positive energy mode parametrized by $\Omega = \Omega_{+}(K)$. Curve 2 is a negative energy mode, while curve 3 is a positive energy mode. Both curves 2 and 3 are described by $\Omega = \Omega_{-}(K)$, see Eqs. (\ref{e2}-\ref{e3}).}
\label{figure3}
\end{center}
\end{figure*}

 As apparent from Fig. (3), it is not possible to have an exact coupling between the negative and positive energy waves described by curves 2 and 3 respectively, since they have no intersection. Nevertheless, from the structure of the graphics, one might suspect that some approximate coupling can take place. To give support to this conjecture, a useful approach is to identify approximate negative and positive energy modes corresponding to the exact waves in an appropriate limit (e.g. for sufficiently large wave-numbers). Hence we follow the style of Ref. \cite{LashmoreDavies} and write the dispersion relation under the factorized form
\begin{eqnarray}
\label{e9}
\Bigl[\Omega &-& K - \frac{1}{\sqrt{2}}\,\left(1+\frac{H^2 K^4}{2}\right)^{1/2}\Bigr]\,\Bigl[\Omega - K + \frac{1}{\sqrt{2}}\,\left(1+\frac{H^2 K^4}{2}\right)^{1/2}\Bigr] \\
&\times& \Bigl[\Omega + K - \frac{1}{\sqrt{2}}\,\left(1+\frac{H^2 K^4}{2}\right)^{1/2}\Bigr]\,\Bigl[\Omega + K + \frac{1}{\sqrt{2}}\,\left(1+\frac{H^2 K^4}{2}\right)^{1/2}\Bigr] \nonumber = \frac{1}{4},
\end{eqnarray}
where two fast
\begin{eqnarray}
\label{e10a}
\Omega \simeq \Omega_{f,r} \equiv  K + \frac{1}{\sqrt{2}}\,\bigl(1+\frac{H^2 K^4}{2}\bigr)^{1/2} \,,\\
\label{e10b}
\Omega \simeq \Omega_{f,l} \equiv  - K + \frac{1}{\sqrt{2}}\,\bigl(1+\frac{H^2 K^4}{2}\bigr)^{1/2},
\end{eqnarray}
and two slow
\begin{eqnarray}
\label{e11a}
\Omega \simeq \Omega_{s,r} \equiv  K - \frac{1}{\sqrt{2}}\,\bigl(1+\frac{H^2 K^4}{2}\bigr)^{1/2} \,,\\
\label{e11b}
\Omega \simeq \Omega_{s,l} \equiv - K - \frac{1}{\sqrt{2}}\,\bigl(1+\frac{H^2 K^4}{2}\bigr)^{1/2} \,,
\end{eqnarray}
approximate space-charge modes can be identified. The subscripts $r$ and $l$ refer to quantum plasma longitudinal oscillations on the rightward and leftward beams, respectively. Indeed, in the reference frame of the beam moving to the right, the Doppler-shifted linear terms in $K$ would disappear in Eqs. (\ref{e10a}) and (\ref{e11a}). Restoring momentarily physical units, we would then have $\omega^2 = \omega_{b}^2 + \hbar^2\,k^4/(4\,m^2)$, which is the quantum modified Langmuir dispersion relation. Here, $\omega$ is the wave-frequency, $k$ is the wave-number and $\omega_b = \omega_{p}/\sqrt{2}$ is the beam's plasma frequency. Moreover, using Eqs. (\ref{e2}--\ref{e3}) it can be proved that $\Omega_+ \simeq \Omega_{f,r} \simeq K + H\,K^2/2$ and $- \Omega_- \simeq \Omega_{s,r} \simeq K - H\,K^2/2$ for large $K$, so that the fast and slow waves are the asymptotic forms of  exact branches of the dispersion relation. It is in this phase velocity context that the $(f,s)$ subscripts refers to ``fast space-charge wave" or ``slow space-charge wave", when focusing on a particular beam \cite{Briggs}. Similar considerations apply to the leftward beam: $\Omega_- \simeq \Omega_{f,l} \simeq -K + H\,K^2/2$ and $- \Omega_+ \simeq \Omega_{s,l} \simeq - K - H\,K^2/2$ for large $K$.

Calculating the time-averaged energy density $<W_{e}>$ for electrostatic oscillations using Eq. (\ref{e5}), it is directly verified that $\Omega_{f,r}$ and $\Omega_{s,l}$ are positive energy modes for all $K$. On the other hand, $\Omega_{f,l}$ and $\Omega_{s,r}$ can have negative energy content depending on special conditions. As for the classical two-stream instability \cite{Briggs}, instability is expected when a wave on a beam has free energy to drive an instability in the counter propagating beam. Inspecting Eqs. (\ref{e10a}--\ref{e11b}) shows that the only possible couplings are $\Omega_{f,l} = \Omega_{s,r}$ or, reversing directions, $\Omega_{f,r} = \Omega_{s,l}$. Focusing on $K > 0$ and positive frequencies, instability is expected when the fast positive energy wave of the rightward beam couples to a negative energy wave on the leftward beam. It is found that the slow wave $\Omega_{s,r}$ is a negative energy mode provided $K < K_C$ (assuming non-negative $\Omega_{s,r}$). Hence the matching condition is $\Omega_{f,l} = \Omega_{s,r}$, or
\begin{equation}
\label{e12}
\frac{1}{\sqrt{2}}\,\left(1+\frac{H^2\,K^4}{2}\right)^{1/2} = K \,,
\end{equation}
as illustrated in Fig. 4, which shows intersection of the fast leftward and slow rightward modes for $H = 0.1$. As in the classical case, the coupling occurs for ${\rm Re}(\Omega) = 0$, in analogy with ideal MHD instabilities \cite{LashmoreDavies}.

\begin{figure*}
\begin{center}
\includegraphics[width=0.7\textwidth]{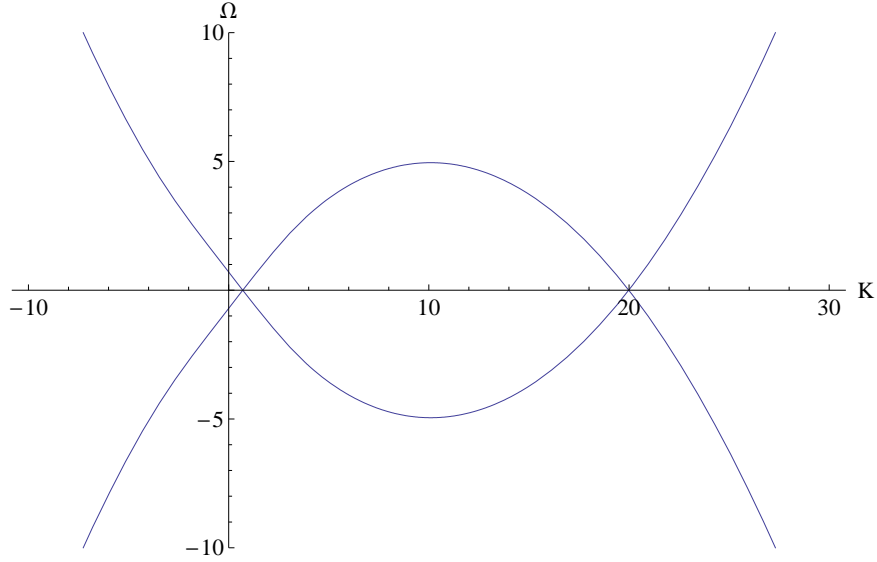}
\caption{Fast leftward $\Omega_{f,l} = - K + (1 + H^2 K^4/2)^{1/2}/\sqrt{2}$ and slow rightward $\Omega_{s,r} = K - (1+H^2 K^4/2)^{1/2}/\sqrt{2}$ asymptotic modes for $H = 0.1$.}
\label{figure4}
\end{center}
\end{figure*}

In the context of this interpretation, the wave-numbers $K_m$ satisfying the coupling condition (\ref{e12}) correspond to maximal instability growth-rate. For $H \neq 0$, they are given by
\begin{equation}
\label{e13}
K^{2}_m = \frac{2}{H^2}\left[1 \pm \sqrt{1-\frac{H^2}{2}}\,\right] \,.
\end{equation}
Notice that in the classical case where $H = 0$,  Eq. (\ref{e12}) yields only the solution $K_{m}^2 = 1/2$.

Taking the plus sign in Eq. (\ref{e13}), one finds
\begin{equation}
\label{e14}
K_m \equiv K_{m,q} = \frac{\sqrt{2}}{H}\left[1 + \sqrt{1-\frac{H^2}{2}}\,\right]^{1/2} \,.
\end{equation}
Assuming $H \ll 1$, we get to leading order,
\begin{equation}
\label{e15}
K_{m,q} = K_C - \frac{H}{8} + O(H^3) > K_B \simeq K_C - \frac{H}{4} + O(H^3) \,,
\end{equation}
where $K_B$ and $K_C$ are defined in Eq. (\ref{e4}). Therefore, $K_B < K_{m,q} < K_C$, which is exactly what should be expected for an instability arising from the coupling of the positive energy mode shown in curve 3 and the negative energy mode shown in curve 2 of Figure 3. This explains the physical origin of the purely quantum instability described in Ref. \cite{Haas},   occurring for large wave-numbers, when $H < 1$. For a fixed value of $H < 1$,  there is a good agreement between $K_{m,q}$  and the exact wave-number for fastest growing quantum instability. The discrepancy comes from the fact that  Eqs. (\ref{e10a}--\ref{e11b}) show just approximate modes.

On the other hand, taking the minus sign in Eq. (\ref{e13}) gives
\begin{equation}
\label{e16}
K_m \equiv K_{m,c} = \frac{\sqrt{2}}{H}\left[1 - \sqrt{1-\frac{H^2}{2}}\,\right]^{1/2} \,.
\end{equation}
We get, to leading order,
\begin{equation}
\label{e17}
K_{m,c} = \frac{1}{\sqrt{2}} + \frac{H^2}{16\sqrt{2}} + O(H^4) < K_A \simeq 1 + \frac{H^2}{8} + O(H^4) \,,
\end{equation}
where $K_A$ is defined in Eq. (\ref{e4}). The wave-number $K_{m,c}$ can be interpreted as the semiclassical branch, since it corresponds to the exact classical wave-number for maximal instability, $K_c = 1/\sqrt{2}$. Moreover $K_{m,c} < K_A$ corresponds to the coupling of the positive (curve 1) and negative energy (curve 2) branches in Figure 3. Finally, we found a satisfactory agreement between the exact and approximate values of the wave-number for maximal growth-rate, at a fixed $H$.

When $H \geq 1$, the elliptic-like branch of Figure 3 disappears. The instability for $K < K_C$ and $H \geq 1$ can also be  understood in terms of negative energy modes. However, we omit these considerations since these parameter ranges are outside the scope of the quantum two-stream model. Indeed, as shown in the next Section, $H$ should be small to avoid the strong coupling regime.

\section{Necessary validity condition for the cold quantum two-stream model}
In a cold two-stream plasma model, be classical or quantum, it is implicit that each beam has a well-defined identity, which is reasonable only if the respective velocity dispersions are much smaller than the beam velocities. It is relevant to analyze the implications of this point in view of  the collisionless hypothesis underlying the original equations \cite{Haas}.

Consider first the case of classical statistics, which applies for sufficiently small particle densities, or large temperatures. In this case, the beam velocity dispersion is the thermal velocity $v_{th}$, so that the present model assumes $u_0 \gg v_{th}$, which is equivalent to a cold beam approximation. Moreover,  the thermal velocity of the beams is a free parameter, independent of the particle density. It can be verified that $g_C = e^2 n_{0}^{1/3}/(\varepsilon_{0}\,m\,v_{th}^2) \ll 1$ (small classical coupling constant), together with $H \geq 1$, implies some relativistic velocities which are out of the scope of the model. Therefore,  $H\ll 1$ is necessary to support the collisionless hypothesis.

Let us now rule out the possibility of a large quantum parameters for dense degenerate plasmas a well. In this case, due to Pauli blocking, the system is all the more ideal that the density is large, and it is appropriate \cite{Landau, Manfredi} to define the quantum coupling parameter  $g_Q = e^2 n_{0}^{1/3}/(\varepsilon_{0}\,T_{F}) \sim (\hbar\,\omega_{p}/(m\,u_{F}^2))^2$. Here, $T_F$ and $u_F$ are the electronic Fermi temperature and velocity, respectively. In the degenerate case, $u_F$ measures the velocity spread of each beam, and increases with the density. Hence, the quantum two-stream model assumes $u_0 \gg u_F$, imposing a bound on the electron particle densities. However, for such small densities one has $H \ll \hbar\,\omega_{p}/(m\,u_{F}^2) \sim g_{Q}^2 \ll 1$. Here again, the collisionless and the well-defined two-stream hypothesis, demand $H\ll 1$.  The present work provide physical arguments to understand the quantum instability branch for small $H$ and large wave-numbers, where the weak coupling regime apply. In addition, large quantum parameters are admissible in the context of a quantum Dawson $N-$stream model \cite{Haas, Dawson}, in which case the beam velocities need not to be large in comparison to the Fermi velocities.

\section{Conclusion}
The main results of this work stem from the factorized form of the dispersion relation, as expressed in Eq. (\ref{e9}). This allows to identify fast and slow waves propagating in both the positive and the negative directions.  The quantum two-stream instability grows when the free energy available in a positive energy wave carried by one beam, is transferred  to a negative energy wave carried by the other beam. The coupling of these waves and their energy exchange has been discussed in terms of their electrostatic energy density, giving rise to stable or unstable scenarios. While the mathematical techniques for the quantum plasma problem are the same as for the classical two-stream plasma, the quantum dispersion relation is more subtle as evidenced by Fig. (\ref{figure3}). The above analysis can in principle be  pursued on similar problems such as the quantum beam-plasma instability and on theories with a discrete structure such as the quantum multi-stream model.

\vskip .5cm
\noindent
{\bf Acknowledgments}\\
This research was financially supported by the Alexander von Humboldt Foundation, by projects FIS 2006-05389 of
the Spanish Ministerio de Educaci\'{o}n y Ciencia and PAI-05-045 of the Consejer\'{\i}a de Educaci\'{o}n y Ciencia de la Junta de Comunidades
de Castilla-La Mancha.  We acknowledge
the anonymous referee for his constructive questions, in particular for pointing out the model's applicability
condition on the parameter $H$.

\end{document}